\documentclass[%
preprint,
 amsmath,amssymb,
 aps,
]{revtex4-2}

\usepackage{graphicx}
\usepackage{dcolumn}
\usepackage{bm}

\usepackage{orcidlink}

\begin{document}


\title{Quantitative, Data-driven Network Model for Global Cascading Financial Failure}

\author{Łukasz G. Gajewski \orcidlink{0000-0003-3097-0131}}%
 \email{lukasz@allfed.info}
\affiliation{Alliance to Feed the Earth in Disasters (ALLFED), 603 S. Public Rd \#57 Lafayette, CO 80026, USA}
\author{Michael Hinge}
\affiliation{Alliance to Feed the Earth in Disasters (ALLFED), 603 S. Public Rd \#57 Lafayette, CO 80026, USA}
\author{David Denkenberger \orcidlink{0000-0002-6773-6405}}
\affiliation{Alliance to Feed the Earth in Disasters (ALLFED), 603 S. Public Rd \#57 Lafayette, CO 80026, USA}
\affiliation{Department of Mechanical Engineering, University of Canterbury, Christchurch, Canterbury 8041, NZ}


\begin{abstract}
Global catastrophic risk events, such as nuclear war, pose a severe threat to the stability of international financial systems. 
As evidenced by even less severe scenarios like the Great Recession, an economic failure can propagate through the world trade network, wreaking havoc on the global economy. 
While the contemporary literature on cascading failure models addresses this issue qualitatively, a simple and intuitive quantitative estimation that could be used in integrated assessment frameworks is missing. 
In this study, we introduce a quantitative network model of global financial cascading failure. 
Our proposal is a fast, efficient, single free parameter model, following a straightforward logic of propagating failures. 
We fit the model to the Great Recession and test it against historical examples and commercial analysis. 
We also provide predictions for a hypothetical armed conflict between India and Pakistan. 
Our aim is to introduce a quantitative approach that could inform policy decisions by contextualising global catastrophic scenarios regarding financial losses and assessing the effectiveness of resilience strategies, complementing existing models and frameworks for broader risk assessment.
\end{abstract}

\maketitle


\newpage
\section{\label{sec:intro}Introduction}
The interconnectedness of the global economy has reached unprecedented levels, driven by advancements in trade, technology, and financial systems. 
While this interdependence has facilitated economic growth and resilience in many contexts, it has also heightened systemic vulnerabilities to Global Catastrophic Risks (or GCRs), which have the potential to collapse civilisation, affecting countless future generations \cite{helbing2013globally, bostrom2011global, centeno2015emergence}. 
Events such as pandemics \cite{rothwell2024global}, financial shocks and geopolitical conflicts have demonstrated the potential to disrupt economies at regional and global scales \cite{wef}. 
The global economy is also vulnerable to additional GCRs without a recent precedent, such as severe volcanic eruptions \cite{newhall2018anticipating, rampino1992volcanic}, nuclear conflict \cite{coupe2019nuclear}, and extreme climate phenomena \cite{world2012thai}. 
Even when originating in a single location, these shocks would likely propagate through supply chains, financial markets, and institutional frameworks, resulting in cascading effects globally \cite{kang2024potential, contreras2014propagation, wang2018simulation, watts2002simple}.

We believe it is vital to have reasonable estimates of the consequences of global catastrophic events to prepare for the case where prevention fails. 
This manuscript aims to introduce a relatively straightforward yet quantitative model in GCR scenarios, not to debate which scenarios are more or less likely to happen; therefore, we will provide a wide range of possible initial conditions in a hypothetical India-Pakistan armed conflict.

We do not intend to compete with detailed economic analysis tools and models \cite{aguiar2019gtap, corong2017standard, boeck2022bgvar} but provide a simple, intuitive and fast way of estimating the order of magnitude of global financial losses in a GCR scenario using publicly available and easily accessible data.
Typical economic models require the modeller to consider many different, complicated, interacting mechanisms, which presents challenges when considering severe market disruptions. 
For example, Computable General Equilibrium Models consider all markets - including all investment, saving and consumer expectations. 
These are powerful tools, but they can be “black boxes” that struggle to link causes to effects and elude interpretability \cite{wing2004computable}. 
They also rely on many buried empirical relationships, which may or may not be valid in more extreme shocks. 
Partial equilibrium models address some of these issues by focusing on the dynamics of a subset of markets; however, this approach also has limitations - chiefly that they will only produce results for a subset of the economic shock. 
In addition, they can also miss important dynamics if unexpected interactions with sectors are excluded from the model \cite{rike2022different, farrow2018welfare}. 
Such models also require much more precise work to formulate and create an intricate web of assumptions that need verification and validation in isolation and as an integral system. While we acknowledge the power of such models and that they constitute the most precise approach today, we also point out that this precision comes at a cost – as outlined above.

Our model is a complex network approach. 
This abstraction allows for a much easier adaptation of the model to different scenarios and enables us to circumvent the intricacies of economic modelling. 
Therefore, a mere fraction of the assumptions is required to produce a result. 
It is a trade-off, of course – we lose the fine detail but retain the big picture.

We propose a singular, intuitive way of how a crisis in one country transfers over to others.
The reason behind this mechanism is most assuredly the result of all the elements that economic models consider, but we opt to abstract all of them and put them under an umbrella mechanism. 
This modelling strategy is employed broadly in many areas of science, such as computational social sciences \cite{albert2002statistical, castellano2009statistical, gajewski2022transitions} and epidemiology \cite{wang2016statistical, pastor2015epidemic, gajewski2022comparison}. 
Using the latter as an example to demonstrate our point here – we recognise that the reason for infection transmission is the complicated fluid and aerosol dynamics and virus/bacteria (micro-)biology, and all other various kinds of complicated and detailed mechanisms, but in terms of modelling infections and their spreading dynamics through a population, we can abstract all of that into a handful of parameters, typically probability of infection and rate of recovery. 
The low computational complexity of our model, O(V+E), where V is the number of vertices and E is the number of edges in the system, additionally makes our approach very suitable for integrated assessment frameworks, similar in spirit to other simplified models in economics \cite{schewe2017role, falkendal2021grain}.

The work presented here can also be considered a quantitative companion to the cascading threshold model \cite{watts2002simple}, which has been used qualitatively to analyse many systems, including supply chains \cite{wang2018simulation} and trade networks \cite{kang2024potential}. 
In our analysis, we assume that the critical value of the threshold parameter has been reached and a system-wide cascading failure is occurring.

\begin{figure}
    \centering
    \includegraphics[width=\linewidth]{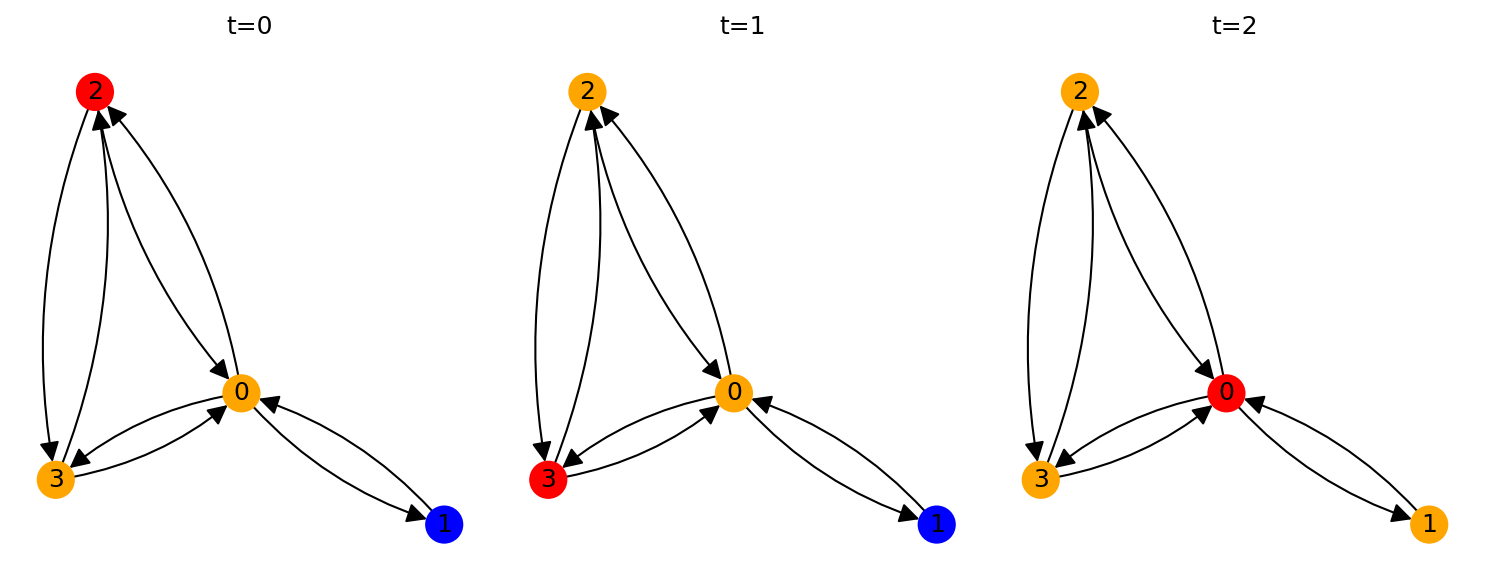}
    \caption{Example propagation. At each time step, the red node experiences a failure and transfers it onto its neighbours (orange) whilst other nodes remain unaffected yet (blue).}
    \label{fig:propexample}
\end{figure}

Our model establishes a directed, weighted graph representing the world trade network, where edges represent import/export between countries (vertices), which in turn have an associated “capacity” – their gross domestic product (GDP) plus the absolute value of net exports. 
A country experiencing a failure, i.e., a fractional reduction $\xi$ of its capacity, transfers that failure onto its neighbours with whom it trades at value $\nu$ via a transfer function $T(\xi; \alpha)$ and reduces this trade value by $\xi$ ($\alpha$ is the sole free parameter of the model). 
Those neighbours then experience their own failure $\xi' = \nu \times T(\xi; \alpha)$ and transfer it onto their neighbours, and so on until all nodes have been affected. 
This is similar to the independent cascade model \cite{kempe2003maximizing}, although our model is deterministic, and we allow countries to hit back the source of their failure. 
The latter, an echo or reverberation effect of sorts, can only happen once between a pair of nodes and at reduced trade volume. 
We present a simple example of such propagation in Fig.~\ref{fig:propexample}, and a more detailed description is available in section~\ref{sec:meth}.

\section{\label{sec:res}Results}
We fit our model to the 2007-2009 Great Recession crisis. 
The Great Depression or World War II would have been better cases; however, data availability prohibits this. 
In the model, we assume a global cascading loss, i.e., all countries are expected to be negatively affected, which was not the case in the Great Recession. 
Thus, we only consider the losses during the 2007-2009 period and ignore the countries that experienced any gains.
In order to provide a prediction interval as opposed to a point prediction, we utilise the quantile regression \cite{koenker2001quantile}. 
The parameter values – $\alpha$ – for specific quantiles found for the Great Recession are then used in all other simulations. 
This approach of combining the quantile regression with fitting to \textit{only} losses in the Great Recession implicitly allows us to capture resilience (or lack thereof) mechanisms on the global market. Fitting to a low quantile level, we essentially say that all countries in our model shall be as resilient as the least affected countries in the Great Recession. Conversely, when we fit to a high quantile level, we say that all countries are vulnerable to financial perturbations. Thus, we achieve a range of estimates for best- and worst-case possibilities in potential crises.

\begin{figure}
    \centering
    \includegraphics[width=\linewidth]{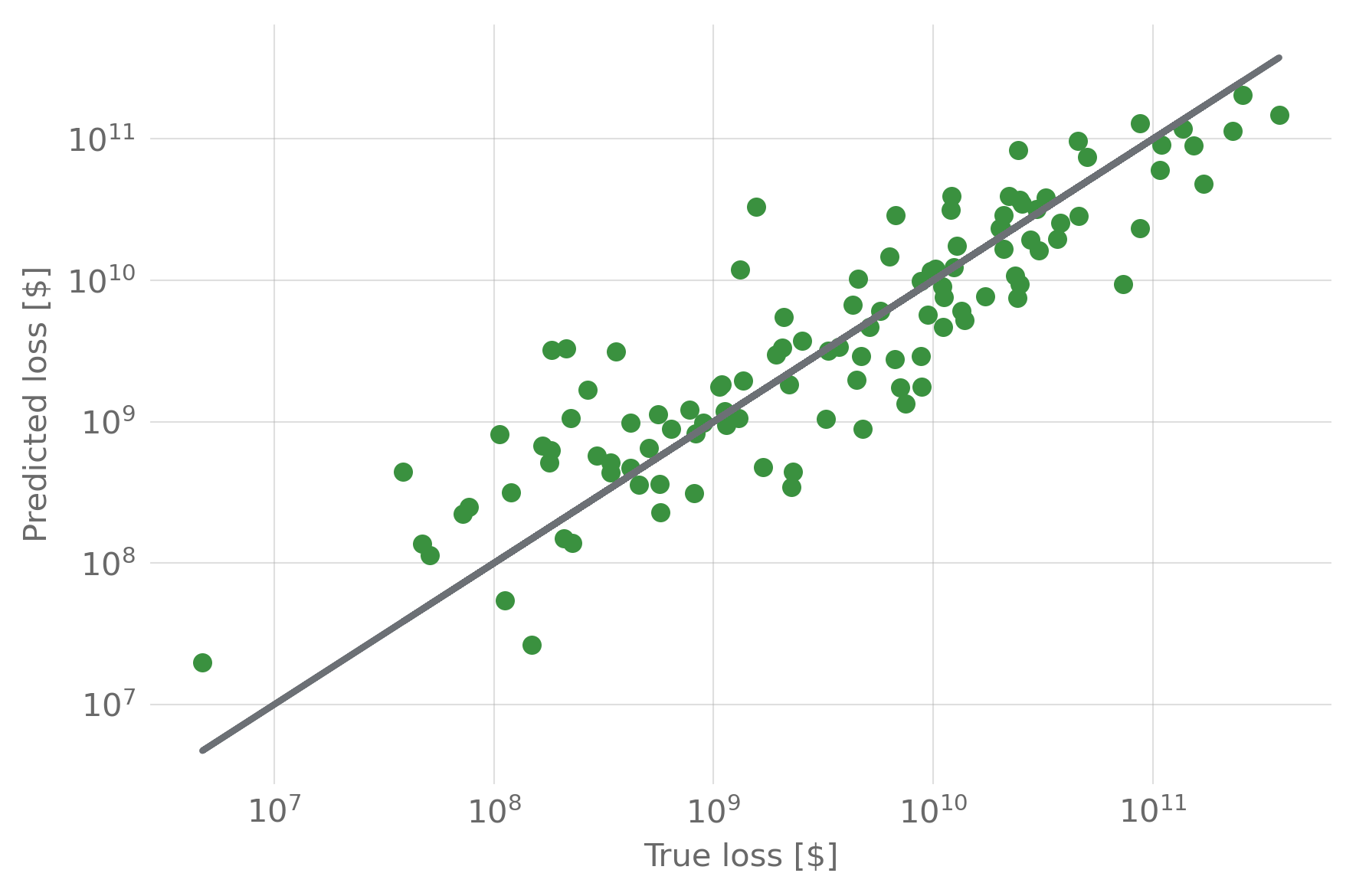}
    \caption{Model financial loss predictions (Y-axis) vs actual losses (X-axis) in US dollars in the Great Recession. The solid line is a visual guide for $Y=X$. In a perfect fit, all points would be on the solid line. The coefficient of determination is $R^2 = 0.66$ (calculated on a linear scale; calculating it in log-log would inflate it to $R^2=0.79$). Since the USA’s loss is the initial condition for this scenario, its losses have been excluded here.}
    \label{fig:fit}
\end{figure}

The comparison of the median predicted losses to actual losses is shown in Fig.~\ref{fig:fit}. 
The true losses were estimated by extrapolating countries’ GDP from the time prior to the crisis to post-crisis and comparing it against the actual GDP value (inflation-adjusted) in that year.

The model predictions here are for the 50th percentile, i.e., the median, and they line up with real values reasonably well with $R^2 = 0.66$. The country-by-country breakdown of the results is presented in Supplementary Information (SI).

\begin{figure}
    \centering
    \includegraphics[width=\linewidth]{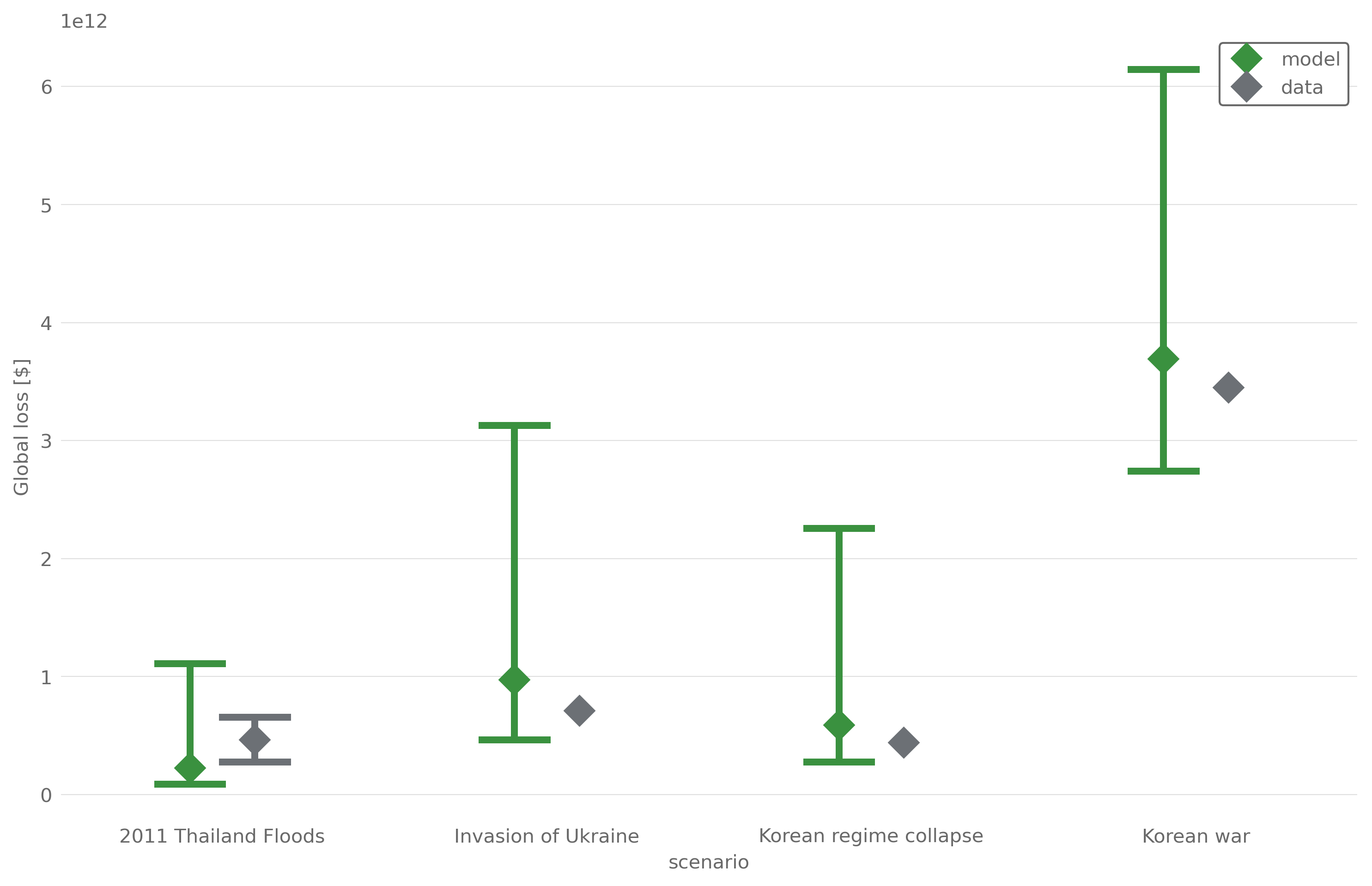}
    \caption{Model predictions vs. our estimates of losses in historical scenarios and the Bloomberg analysis of a potential Korean conflict. The Y-axis shows the financial losses in trillions of US dollars. Green intervals are the output of our model, while grey are real-value estimates (i.e., target values). The error bars are 25th to 75th percentiles.}
    \label{fig:predict}
\end{figure}

The validation of our model is, of course, a challenge due to, fortunately, very few recent global disasters that would be appropriate. 
However, we assembled a set of historical examples alongside the analysis of the Korean peninsula tensions by Bloomberg Economics \cite{bloomberg}, and the predictions of our model are shown in Fig.~\ref{fig:predict}. 
The prediction interval chosen here was $50\%$, meaning that the lower estimate is with the optimal parameter value for the 25th percentile and the upper estimate for the 75th percentile, with the diamond shape within the range representing the median. 
While the model matches the real value estimates for this interval, it is important to remember the model's limitations here. In scenarios like the 2014 Russian financial crisis, the model would not be successful. This is because the model lacks a mechanism for gaining or not losing money. The 2014 case is beyond the current capabilities of the model as it was not a global cascading failure \cite{wsj, viktorov20202014, cnn}.

\begin{figure}
    \centering
    \includegraphics[width=\linewidth]{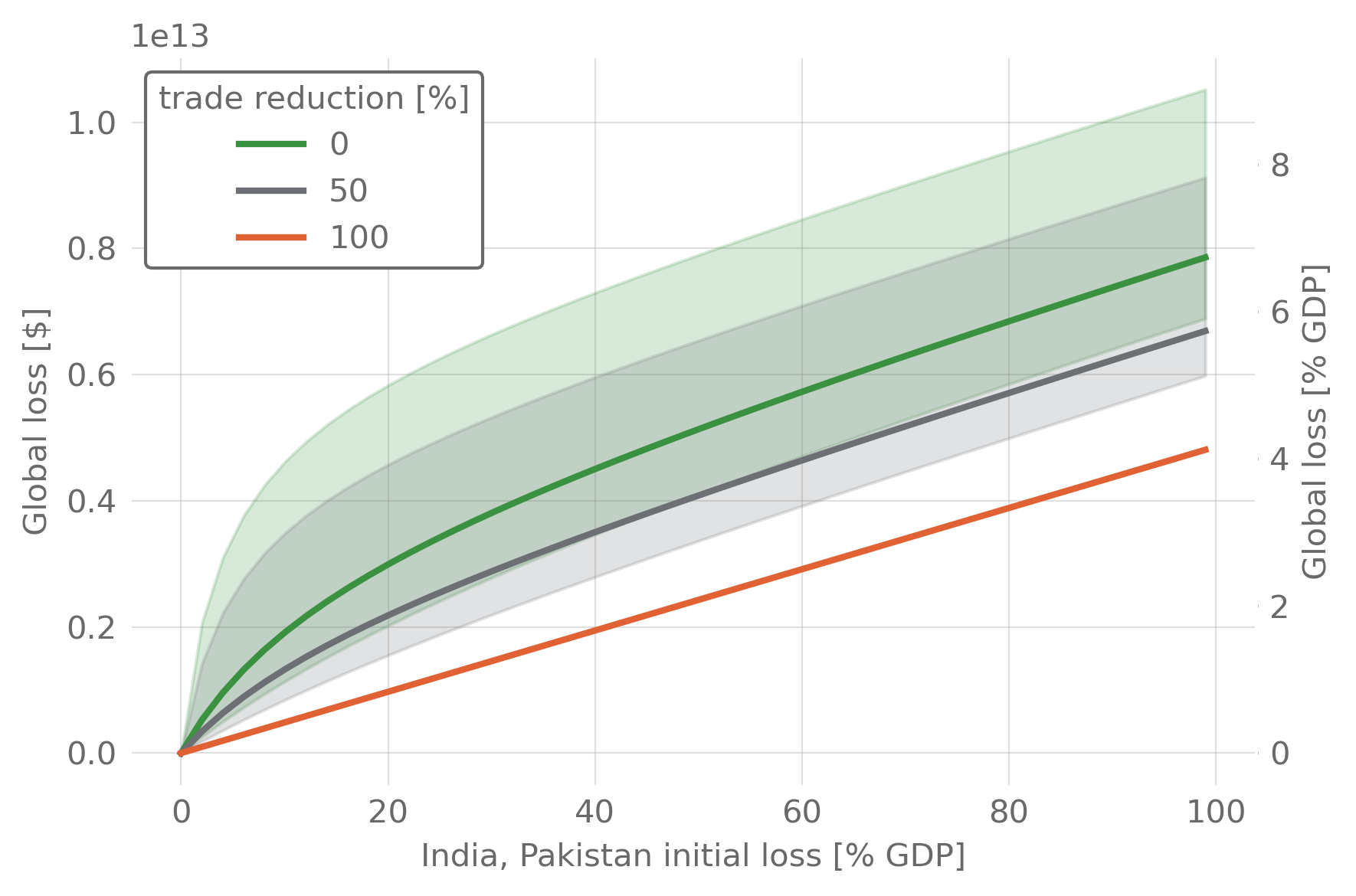}
    \caption{India-Pakistan conflict predictions. On the X-axis, we show the \% of GDP loss of each India and Pakistan, and on the Y-axis, the total global loss (in tens of trillions of dollars on the left axis and \% of global GDP on the right). Results are $50\%$ prediction intervals except for the orange line, in which the only losses are India’s and Pakistan’s. Colours indicate the amount of trade headed to India and Pakistan that has been diverted to other countries. These results as the \% of global GDP loss on both axes are shown in SI.}
    \label{fig:india-pakistan}
\end{figure}

With these results, we are confident that our model can accurately predict the magnitude of global financial losses in a catastrophic global scenario, and we can now consider a new scenario of the hypothetical India-Pakistan armed conflict to depict the type of predictions our model is capable of.

Fig.~\ref{fig:india-pakistan} depicts the median prediction (solid line, green, $0\%$ trade reduction) with the $50\%$ prediction interval as the band around it. 
The total loss is calculated as a function of the initial loss inflicted upon India and Pakistan due to their conflict, such that an x-axis value of, for example, $20\%$ means that India and Pakistan each lose $20\%$ of their corresponding capacities. 
For the sake of concrete comparison, the Bloomberg analysis indicates that South Korea would lose $37.5\%$ of its GDP due to war \cite{bloomberg}, so assuming a $37.5\%$ loss for each country ($\$1.74$ trillion), the median prediction from our model is a $\$4.4$ trillion loss.
However, this assumption might not apply to India and Pakistan; therefore, we provide our own estimate of the initial conditions of this conflict.

\begin{figure}
    \centering
    \includegraphics[width=\linewidth]{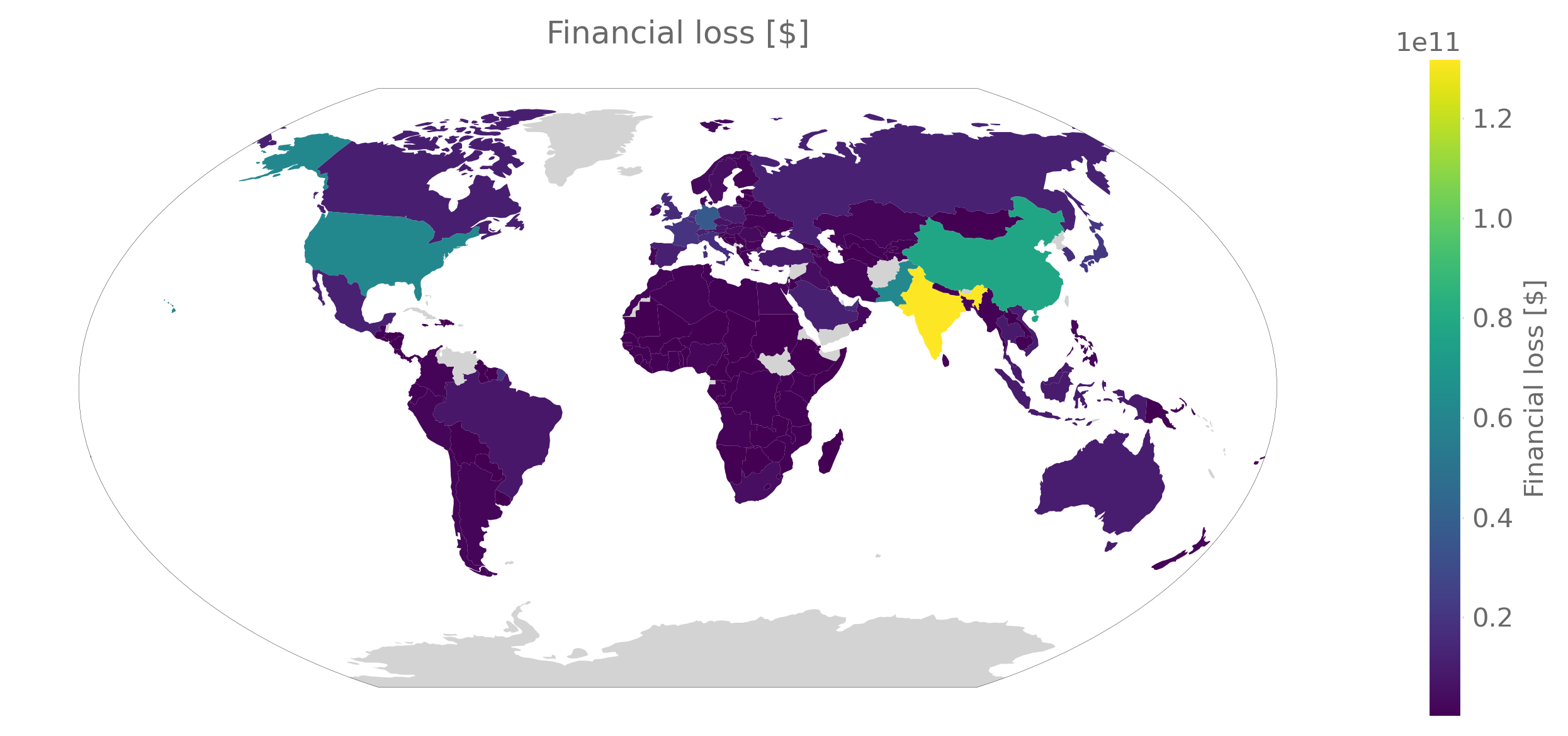}
    \caption{India-Pakistan example scenario maps showing the absolute loss in US dollars. The colour depicts the loss magnitude of model's median prediction. The relative loss map is included in SI.}
    \label{fig:india-pakistan-map}
\end{figure}

We use the industrial GDP and fatalities estimates from \cite{blouin2024global}. 
In that study, the authors estimate the industrial GDP and immediate fatalities due to a nuclear conflict between India and Pakistan. 
The industrial GDP of India would decrease by $1.5\%$ and Pakistan’s by $8\%$, and the countries would lose 33 million and 24 million people, respectively.

Assuming that India’s industrial output is $25\%$ of its GDP and Pakistan's $20.8\%$ \cite{cia-india, cia-pak}, we estimate the initial loss as the sum of the number of fatalities multiplied by GDP per capita and the industrial loss, resulting in India’s initial loss being $2.7\%$ and Pakistan’s $11.6\%$.
With these assumptions, the median prediction for the total global losses is $\$812$ billion.

In Fig.~\ref{fig:india-pakistan}, we also consider a straightforward experiment of how redistributing trade from countries experiencing failure impacts the results; those are the grey line for $50\%$ trade reduction and the orange for $100\%$. 
The idea here is very simple – the specified amount of trade that some country has with India and Pakistan is diverted evenly to said country’s other neighbours. The $100\%$ reduction line doesn’t have a prediction interval because the only losses then are due to India and Pakistan alone.

Finally, in Fig.~\ref{fig:india-pakistan-map}, we show a detailed world map indicating financial losses (median prediction) for each country (as long as we have sufficient data to do so) in the example scenario of $2.7\%$ and $11.6\%$ initial capacity losses in India and Pakistan, respectively.
Naturally, India and Pakistan suffer significantly, and in terms of absolute value, the USA and China are “leading” as well. That is likely due to the fact that these are simply big and well-connected economies in the vein of “the taller they stand the harder they fall”.

\section{\label{sec:dis}Discussion}
In this paper, we propose a quantitative network model of global financial cascading failure. 
Our approach is fast and efficient and can be run on virtually any modern machine without needing cloud or super-computing. 
It is also straightforward, having only one free parameter, and intuitive as, in its essence, it is a “passing on” procedure with a rebound mechanism. 
Our results match more complex procedures and historical examples very well, and while the model might lack some nuance, we believe it can be effective for estimating financial losses in global catastrophic risk scenarios.

In practical terms, we hope that our model can be a helpful tool to inform policy decisions regarding preparedness for catastrophic situations. 
At its core, it allows for contextualising global catastrophic scenarios in terms of financial losses and, in principle, it can help identify the effectiveness of resilience strategies in specific scenarios. 
Some strategies are naturally easier than others to be evaluated by our model, as portrayed by the simple example of trade redistribution. 
Thus, trade diversification or scaling-up of domestic production would be good candidates for policymakers to consider. We also expect it to be a valuable addition to existing models and integrated assessment frameworks \cite{rivers2024food} alongside other mechanisms such as nuclear winter or any other social, economic or political ones. 
It is especially vital to consider cascading effects because, as indicated before \cite{kang2024potential, contreras2014propagation, wang2018simulation, watts2002simple} and supported by our own results, small perturbations to the system can cause a major ripple effect affecting the whole system.

Despite what we believe to be a successful demonstration of the model’s capabilities, the model is not without its flaws and limitations. 
This paper is a proof-of-concept, and while we think that this approach can already be applied to assessing the financial impacts of global catastrophic risk scenarios, there is plenty of room for improvement and consideration. 

Firstly, its structure and primary operational mechanism are conjectured and not derived from first principles (socio-economic or otherwise). 
This, perhaps, is not a fatal flaw on its own; however, it is easy to imagine increased accuracy and precision from a derived transfer function rather than a conjectured one. 
We also suppose a single, independent cascade dynamic while others, more akin to compartmental models, such as Susceptible-Infected-Removed, Susceptible-Infected-Susceptible or reaction–diffusion \cite{scagliarini2025assessing}, could be considered.
Secondly, as it stands now, it has no allowance for not losing or gaining capital. Lack of loss could be potentially addressed by directly combining our approach with the threshold dynamics studied, e.g., in \cite{kang2024potential} or similar. 
This poses its own challenges; however, finding a way of estimating the real-world value of the control parameter is the major one. 
Financial gains are severely more complex as they involve certain countries taking over trade connections, which could be addressed with temporal networks \cite{holme2012temporal, chung2020spatial}.
Thirdly, it heavily relies on historical data to find optimal parameter values. 
Due to the lack of high-quality data on a complete global cascading failure, the straightforward loss estimate and fitting procedure we employed are naturally flawed. 
Since we hope that an appropriate training scenario does not happen any time soon, this can prove to be a challenging limitation to remove.
Lastly, the initial conditions must be estimated via some other means, which might differ significantly between different catastrophic scenarios, although some of that work has already been done, as in the estimates we have taken from \cite{blouin2024global}. 

\section{\label{sec:meth}Methods}
All the data used in our study are publicly available from the International Monetary Fund (IMF) \cite{imfdata} and World Bank \cite{worldbankdata} data portals. 
The model is implemented in Python \cite{python} using complex networks and data science libraries \cite{SciPyProceedings_11, mckinney-proc-scipy-2010, harris2020array, reback2020pandas, Hunter:2007, Waskom2021, 2020SciPy-NMeth}.
The model's implementation is open-source and available at \url{https://github.com/allfed/cascading-financial-failure}.

\subsection{Data Processing}
The trading data from the IMF data portal were chosen to be the CIF (Cost, Insurance, and Freight) import data following the logic in \cite{kang2024potential} that presumes this is the most reliable option available.
All values are presented on a nominal US dollar (USD) basis relevant to the year in which the shock occurred, with the World Bank data acquired in constant 2015 USD and then adjusted for inflation using US CPI, also taken from the World Bank database.

The data used in the analysis is from the appropriate year depending on the scenario analysed, i.e., 2007 for the Great Recession, 2018 for the Korean crisis (this year was chosen for better comparison with the Bloomberg result), 2023 for the India-Pakistan war, etc. 
The list of countries used in each scenario depends on the data availability in the given year. When certain countries lack appropriate data, they are removed from the dataset for that scenario.

To estimate the financial loss in terms of GDP in the 2007-2009 recession, we use linear regression fitted to the four years prior to the event (so in this case, years 2004-2007), then extrapolate to the expected GDP in the year 2009, from which we subtract the actual value in 2009, giving us the loss value in USD. 
The idea here of using a local linear approximation is rooted in what is typically used in assessing losses by, e.g., the World Bank \cite{world2012thai}, i.e., extrapolating from one year to the next and comparing against the actual value. 
GDP as a series in long timespans (e.g. 50 years) are often non-linear; however, they are also typically not available, so a long-term trend is not feasible. 
In “medium” timespans, there is not even a guarantee of a monotonic series (e.g., Russia’s GDP time series in the years 1990-2007 is parabola-esque). 
On the other hand, locally, i.e., in short timespans, the time series can often be considered linear with reasonable accuracy, while very short, i.e., two years, period risks fitting to outlier events.
More sophisticated and country-tailored approaches would naturally be better. 
Still, they fall outside this paper's scope as our primary goal here is to propose a quantitative cascading network model. 
While we believe this is an appropriate approach for our purposes, we show how the model performs with other timespans in SI.

\subsection{Validation scenarios}
In the invasion of Ukraine, the initial conditions – the loss for Ukraine and Russia – are calculated using the linear extrapolation technique, the same as for the Great Recession. 
The global losses, however, are taken from the World Bank analysis \cite{world20222ukraine}, which estimates them to be $0.7\%$ of global GDP.

In the 2011 Thailand floods, once again, initial conditions are estimated like in previous cases, with the exception that the four-year period before 2011 happened to overlap with the Great Recession, which is a significant outlier event. 
Thus, we consider the years $\{2006, 2007, 2010\}$ here. 
The global losses, however, are again estimated based on the World Bank report \cite{world2012thai}, in which there is an estimate for the damage to the global industrial output of $2.5\%$. 
To convert it to financial losses, we looked at the share of GDP by sector and country \cite{owid}, and the lowest and highest values in this data are used to create the upper- and lower-end estimates for the global financial losses due to the floods, such that it’s $2.5\% \times 43.35\% \times \text{global GDP}$ and $2.5\% \times 18.4\% \times \text{global GDP}$, respectively.

Finally, the Korean peninsula scenarios are taken from the Bloomberg analysis \cite{bloomberg}.
This analysis considers two hypothetical scenarios of the North Korean regime collapse and an armed conflict between North and South Korea.
While it cannot be considered a fact as this analysis is also a result of a model (or several models), this model in question is far more complex than ours and well-established in economic literature \cite{aguiar2019gtap, corong2017standard, boeck2022bgvar}. 
A keen reader might notice a slight discrepancy between the Bloomberg values depicted in our plots and the original article. 
This difference is due to the different data sources used in their analysis compared to ours. 
To remedy this mismatch, we used the percentage changes from the Bloomberg analysis and calculated the absolute values of the 2018 data that we could obtain.
This is necessary because, in their analysis, the data are extrapolated in an unspecified way to 2024, so we cannot match the datasets. 
North Korea is omitted from our model due to a lack of appropriate data.

\subsection{Model}
The structural part of the model is a world trade network built from the IMF trading data.
We create a directed, weighted graph where the direction indicates import/export, and the weight is the value of the trade in USD. 
Each node has a “capacity” attribute whose value is given by the country’s GDP plus the absolute value of its net exports.

The dynamic part of the model is inspired by the threshold cascading failure model \cite{kang2024potential, watts2002simple, wang2018simulation} and the independent cascade model \cite{kempe2003maximizing}; however, in our setting, we are past the critical threshold value and the evolution is deterministic. 
The reason for this is twofold. 
Firstly, we are most interested in global catastrophic scenarios, and secondly, we aim to isolate the dominant cascading failure mechanism. 
Thus, our model presumes that one country's financial crisis propagates throughout the global trade network like a cascading failure with an addition of reverberation. 
Each country experiencing a failure transfers that loss onto its neighbours, which causes them to experience a failure, which they transfer onto their neighbours and so on.
The reverberation means that a country transferring its failure onto others gets hit back by the failure echo from its neighbours, but this can happen only once per pair of nodes.
Moreover, our approach is quantitative as opposed to the aforementioned threshold models, in the sense that we do not just consider a change of state but a transfer function of one country’s loss onto the next. 
This is in contrast to models that are typically only interested in a binary failed/not-failed state.

The transfer of the loss from one country to another is:
\begin{equation}
    T(\xi; \alpha) = \frac{\alpha\xi}{\alpha+2\xi-1},
\end{equation}

where $\alpha$ is the free parameter of the model, and $\xi$ is the fractional capacity reduction $\xi\in{[0, 1]}$.

This transfer function is derived by setting the mode of the beta distribution to $\xi$ and calculating the expected value of the distribution. 
This distribution is a natural choice for modelling random variables bounded in the $[0, 1]$ domain \cite{gupta2004handbook}. 
On the other hand, this choice is somewhat arbitrary, and in principle, any function $[0, 1] \rightarrow [0, 1]$ could be utilised here. 
We consider two other transfer functions, a linear and a quadratic one, in SI.
When country A transfers its loss $\xi_A$ onto country B, the capacity of country B is reduced by $\nu(A, B) \times T(\xi_A; \alpha)$ where $\nu(A, B)$ is the weight of the edge $(A, B)$, and this weight is also updated such its new value is $\nu(A, B) \times \xi_A$. 
Country A does this to all its neighbouring countries, and then each of those neighbours will get their turn to repeat the process. Each country transfers its loss to its neighbours only once.

The evolution of the cascade follows the breadth-first search (BFS) traversal, reminiscent of agent-based implementations of models such as Susceptible-Infected or Independent-Cascades \cite{gajewski2021detecting} (see Fig.~\ref{fig:propexample} for an example of the propagation). 
Since BFS is the core computational algorithm, our model's theoretical computational complexity is $O(V+E)$ time, where $V$ is the number of vertices and $E$ is the number of edges; thus, running on virtually any modern machine is fast and efficient. 
While this isn’t a major consideration for small systems such as the country-to-country trade network (order of magnitude – $150$ nodes), our model could be applied to other, more granular trading or supply chain systems, and the computational complexity would become of greater importance. 
We find that the order in which the neighbours are "visited" alters the results slightly for the transfer function we have chosen and can be devastating in other cases. We discuss it in detail in SI, but for the purposes of the main text, we order the visitation schedule by neighbours’ capacity in an ascending order.

\subsection{Fitting}
We find the best $\alpha$ that minimises the so-called pinball loss at the specified quantile $q$.
The pinball loss is defined as:
\begin{equation}
    L(y, \hat{y}) = \frac{1}{n}\sum_{i=0}^{n-1} q\max{(y_i-\hat{y_i}, 0)} + (1-q)\max{(\hat{y_i}-y_i, 0)}
\end{equation}

Where $y$ is a vector of real values and $\hat{y}$ of model estimates for $y$.
Therefore we want to find $\alpha$ for our model that produces the vector $\hat{y}$ such that $L(y, \hat{y})$ is at its minimum for a provided quantile $q$.
In our case, due to a large variance in the data, the $y$ vector consists of the logarithm of financial losses for each country, i.e., $y_i$ is the logarithm of the financial loss of country $i$.
This procedure is known as the quantile regression \cite{koenker2001quantile}. 
The idea is to find a fit that best represents a chosen data quantile. This allows us to provide a prediction interval instead of a point. 
The search for optimal $\alpha$ is conducted with the SciPy package \cite{2020SciPy-NMeth} and is always done using the Great Recession data.

\begin{acknowledgments}
We thank Simon Blouin for useful comments on the manuscript.
\end{acknowledgments}

\nocite{*}

\bibliography{ms}

\end{document}



\title{Quantitative, Data-driven Network Model for Global Cascading Financial Failure\\ Supplementary Information}

\author{Łukasz G. Gajewski \orcidlink{0000-0003-3097-0131}}%
 \email{lukasz@allfed.info}
\affiliation{Alliance to Feed the Earth in Disasters (ALLFED), 603 S. Public Rd \#57 Lafayette, CO 80026, USA}
\author{Michael Hinge}
\affiliation{Alliance to Feed the Earth in Disasters (ALLFED), 603 S. Public Rd \#57 Lafayette, CO 80026, USA}
\author{David Denkenberger \orcidlink{0000-0002-6773-6405}}
\affiliation{Alliance to Feed the Earth in Disasters (ALLFED), 603 S. Public Rd \#57 Lafayette, CO 80026, USA}
\affiliation{Department of Mechanical Engineering, University of Canterbury, Christchurch, Canterbury 8041, NZ}

\maketitle

\section{Country by country results}

\begin{figure}[!htb]
    \centering
    \includegraphics[width=\linewidth]{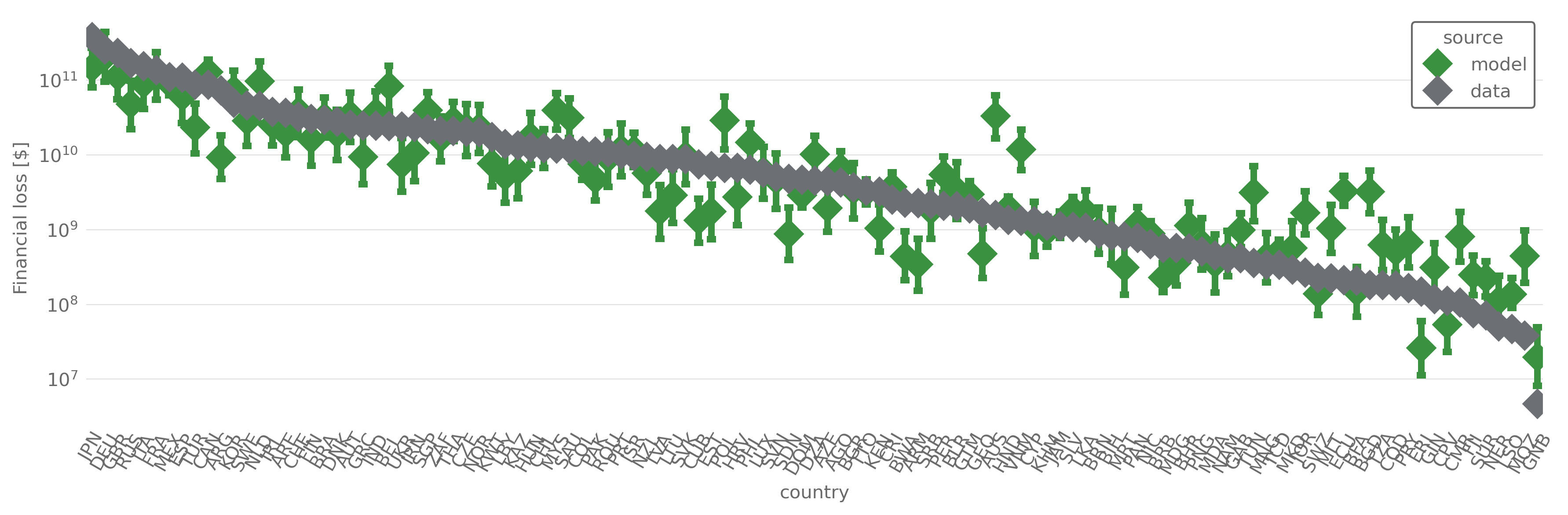}
    \caption{Country-by-country prediction and real-data estimate in the training data (the Great Recession).}
    \label{fig:c_by_c}
\end{figure}

In Fig.~\ref{fig:c_by_c}, we show how the prediction intervals fit the data estimates of GDP losses of the Great Recession. We find that our model’s hit rate (i.e., the number of prediction intervals overlapping with the data points) roughly follows linearly with $\Delta q$, which is the difference between the upper and lower quantiles. 
Detailed results of this relation are shown in the next section.
The tail end of the losses starts exhibiting the effects of “little to no change” discussed in the main text. We omit countries for which we do not have sufficient data or did not experience a loss in our real value estimation.

\section{Alternative hyper-parameters}
\begin{figure}
    \centering
    \includegraphics[width=\linewidth]{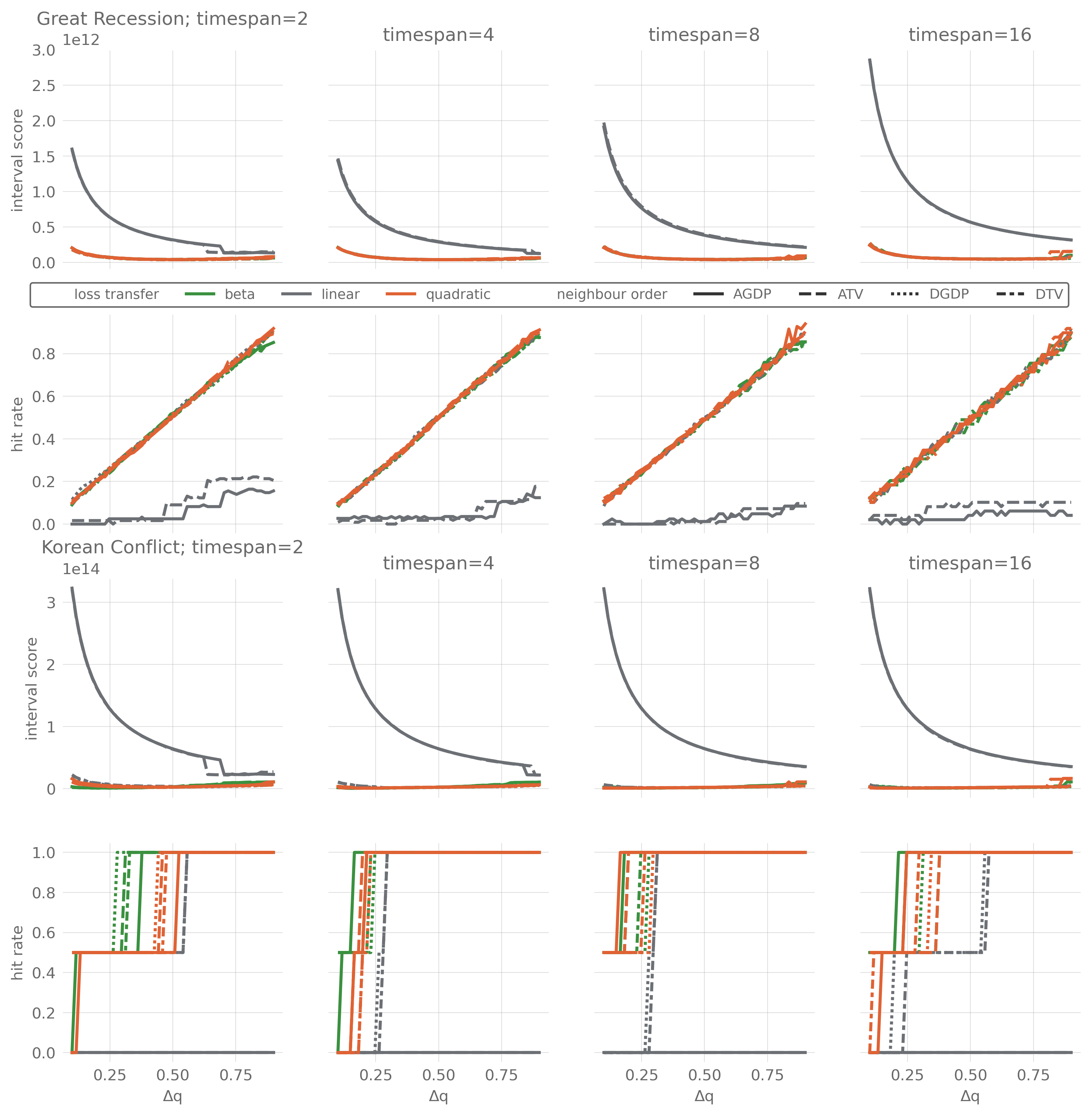}
    \caption{Model performance in various conditions. The first two rows consider the Great Recession fit. The last two rows predict the performance of the Korean conflict using the parameters found with the Great Recession data. Each column corresponds to a different timespan used for the estimation of the real value of losses in the Great Recession (see Methods in the main text). Rows one and three show the interval score values (the lower, the better), while two and four the hit rate (higher is better). Colours indicate the different transfer functions, while line styles the ordering of neighbours visited during the graph traversal.}
    \label{fig:comp_1}
\end{figure}

Here, we discuss our model's hit rate and interval score as evaluated on the Great Recession and the Korean peninsula conflict analysis. We consider the dependence of the results on a) the transfer function, b) the order in which the neighbours are affected, and c) the timespan used for our real value estimates of losses.

The interval score \cite{gneiting2007strictly}, which penalises the model for large prediction intervals, is defined as:

\begin{equation}
    I(y, \hat{y};\Delta q) = \hat{y_u} - \hat{y_l} + \frac{2}{1 - \Delta q}(\hat{y_l}-y)[y < \hat{y_l}] + \frac{2}{1 - \Delta q}(y-\hat{y_u})[y > \hat{y_u}],
\end{equation}

where $y$ is the real value, $\hat{y}$ is the model's predictions. $\hat{y_u}$ and $\hat{y_l}$ correspond to the upper and lower quantile estimate, respectively, and $\Delta q$ is the difference between these quantiles. $[\dots]$ is the Iverson bracket notation \cite{knuth1992two}.

The hit rate is simply the number of prediction intervals which overlap with the target data values.

Aside from the transfer function described in the main text (labelled as “beta” here), we considered two other examples:
Linear, defined as

\begin{equation}
T(\xi; \alpha) = \text{min}(\alpha  \xi, 1),    
\end{equation}

and quadratic:

\begin{equation}
T(\xi; \alpha) = \text{max}(\text{min}(\alpha  \xi^2 + (1-\alpha)\xi, 1), 0).    
\end{equation}

The min and max functions are necessary to secure the mapping $[0, 1] \rightarrow [0, 1]$.

As far as the ordering of neighbours is concerned, we consider four cases: ascending and descending (denoted by the first letter of the label in plots “A” or “D” respectively) by either the node’s capacity (labelled as “GDP”) or the trade value (labelled as “TV”) with the country initiating the transfer. 

\begin{figure}
    \centering
    \includegraphics[width=\linewidth]{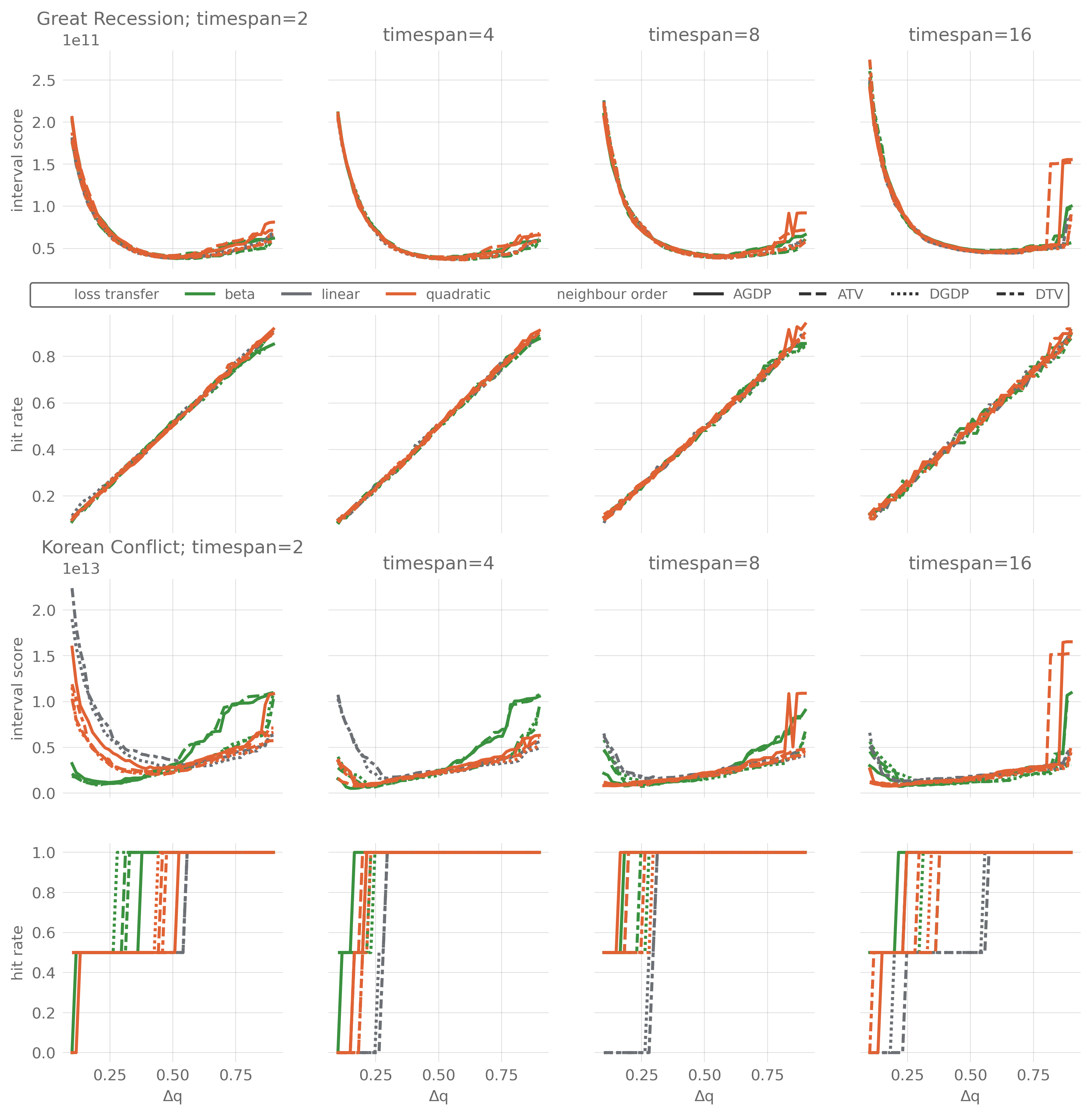}
    \caption{Model performance in various conditions with the worst performing variants removed. The first two rows consider the Great Recession fit. The last two rows predict the performance of the Korean conflict using the parameters found with the Great Recession data. Each column corresponds to a different timespan used for the estimation of the real value of losses in the Great Recession (see Methods in the main text). Rows one and three show the interval score values (the lower, the better), while two and four the hit rate (higher is better). Colours indicate the different transfer functions, while line styles the ordering of neighbours visited during the graph traversal.}
    \label{fig:comp_2}
\end{figure}

In Fig.~\ref{fig:comp_1}, we show a series of panels illustrating all of those dependencies. 
The first four panels are the interval score for the Great Recession fit, and the next four depict the hit rate, both as a function of $\Delta q$. 
Each column corresponds to a different timespan used (in years; four are used in the main text). 
Colours indicate the transfer function while the style of lines the neighbours’ order. The following (third) row shows the interval score calculated for the Bloomberg analysis values, and the last row the hit rate for those values. In these cases, we, as always, use $\alpha$ values found for the Great Recession. 
It is immediately apparent that the linear transfer function is much more reliant on the choice of the ordering than the other two. In the case of both of the ascending options, the linear transfer function completely fails to fit the data. 
Therefore, in Fig.~\ref{fig:comp_2}, we show the same plot, but these two variants (linear and ascending) were removed.
While there is not a definitive winner that is best in all cases, for the timespan of two years, the “beta” variant seems to be most appropriate, with quadratic being a close second, so this is the one used in the main text. We do not argue that the beta transfer function is the best possible, but the investigation of what would be the best function here is beyond the scope of this paper.

\begin{figure}
    \centering
    \includegraphics[width=\linewidth]{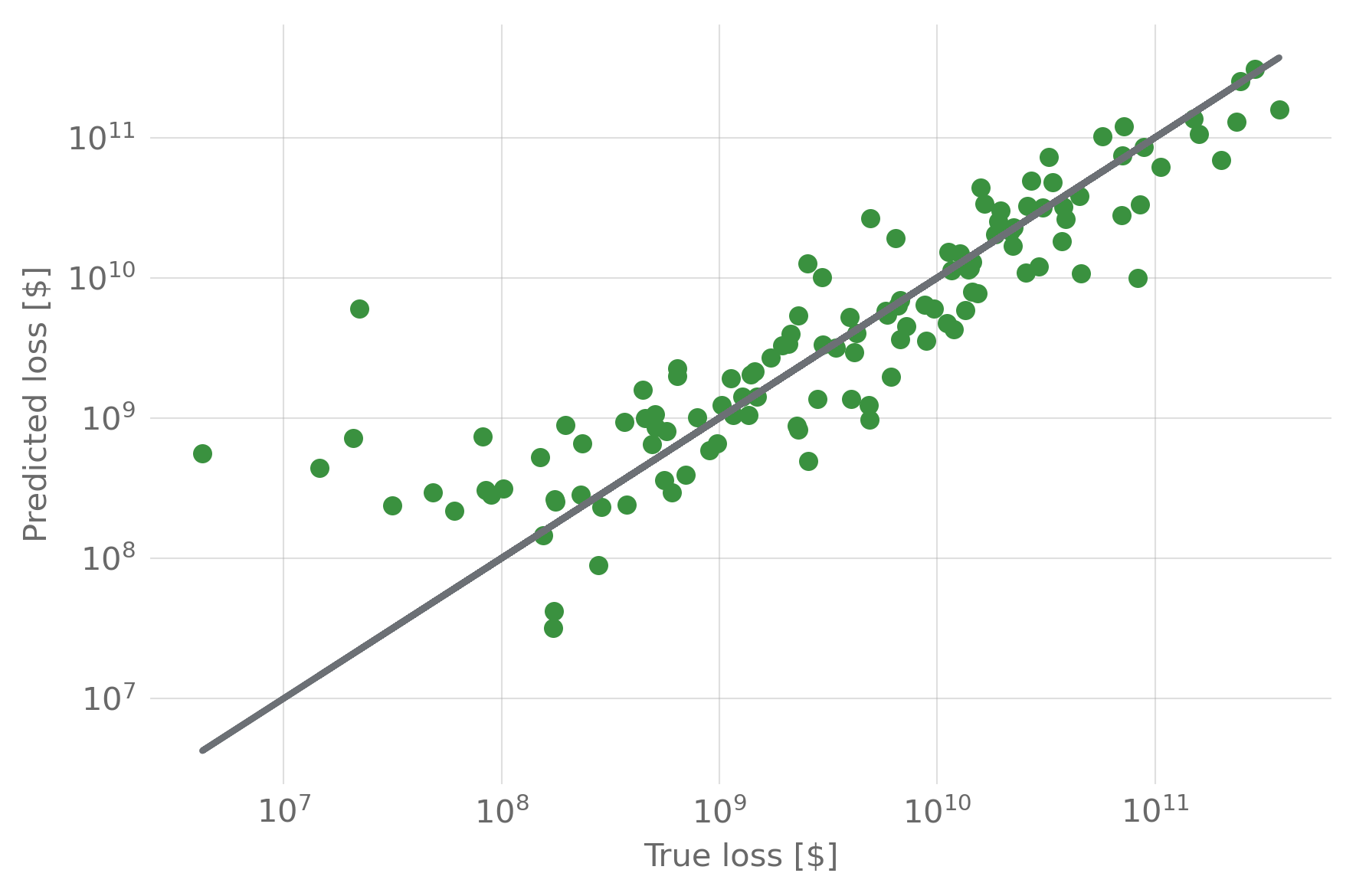}
    \caption{Model financial loss predictions (Y-axis) vs actual losses (X-axis) in US dollars in the Great Recession. The solid line is a visual guide for Y=X. In a perfect fit, all points would be on the solid line. The coefficient of determination is $R^2 = 0.76$ (calculated on a linear scale). Since the USA’s loss is the initial condition for this scenario, its losses have been excluded here. Here, we considered the timespan of two years and the descending by GDP ordering of neighbours.}
    \label{fig:fit_si}
\end{figure}

\begin{figure}
    \centering
    \includegraphics[width=\linewidth]{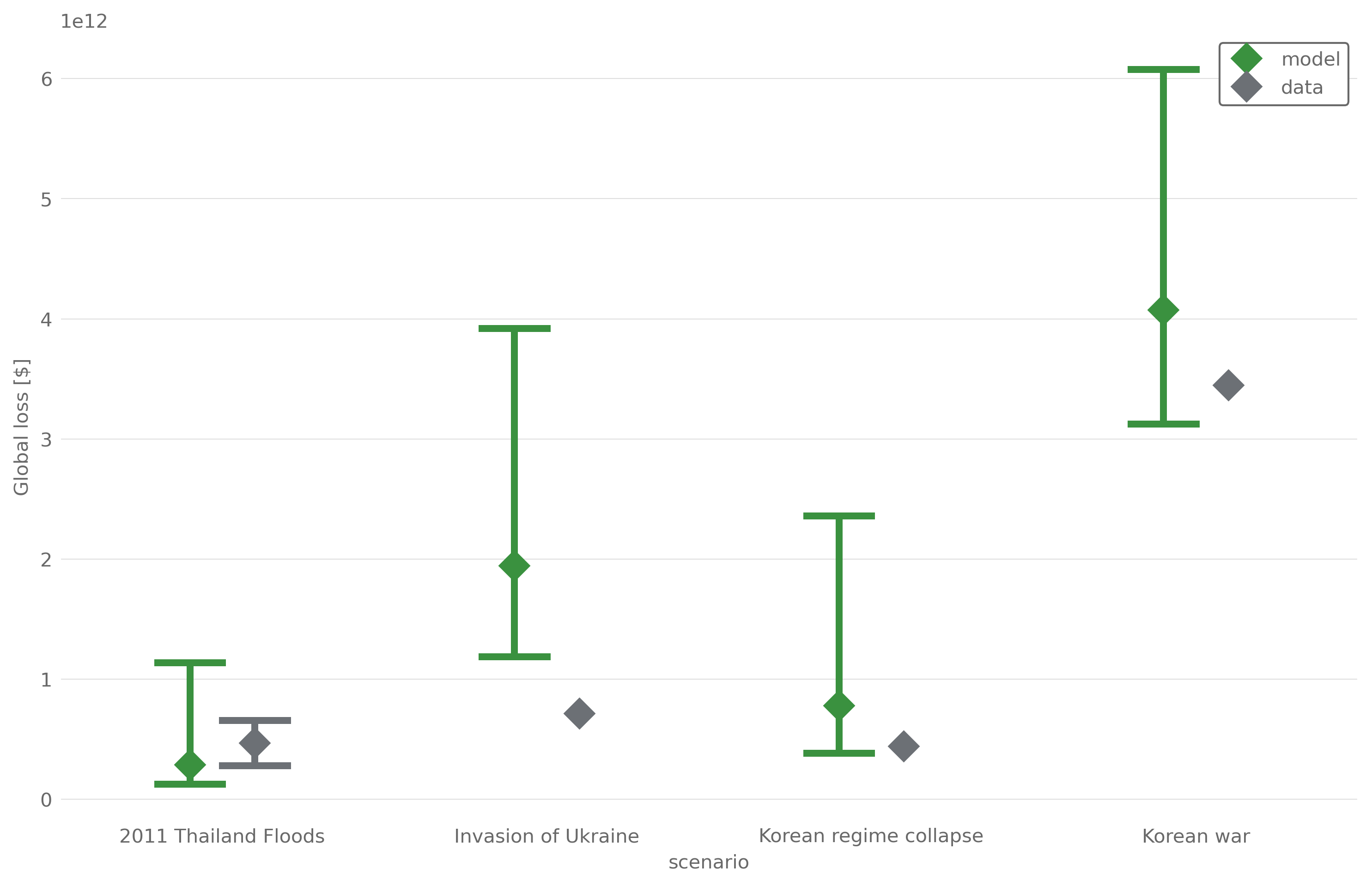}
    \caption{Model predictions vs. our estimates of losses in historical scenarios and the Bloomberg analysis of a potential Korean conflict. The Y-axis shows the financial losses in US dollars. Green intervals are the output of our model, while grey are real-value estimates (i.e., target values). The error bars are 25th to 75th percentiles. Here, we considered the timespan of two years and the descending by GDP ordering of neighbours.}
    \label{fig:other_si}
\end{figure}

To give an example of how all this affects the results shown in the main text, in Fig.~\ref{fig:fit_si} and~\ref{fig:other_si}, we present alternative versions of Fig.~2 and Fig.~3 from the main text. 
Here, we considered a timespan of 2 years for the purposes of fitting the model and visited the neighbours in a descending GDP order. 
The model still performs relatively well; in fact, the $R^2$ for the Great Recession is higher, although the smaller losses are not as well represented (i.e., the model skews more heavily towards large losses), and cumulative losses in some of the validation scenarios require higher $\Delta q$ to fit all the scenarios.

\section{India-Pakistan results in relative metrics}

\begin{figure}[!htb]
    \centering
    \includegraphics[width=\linewidth]{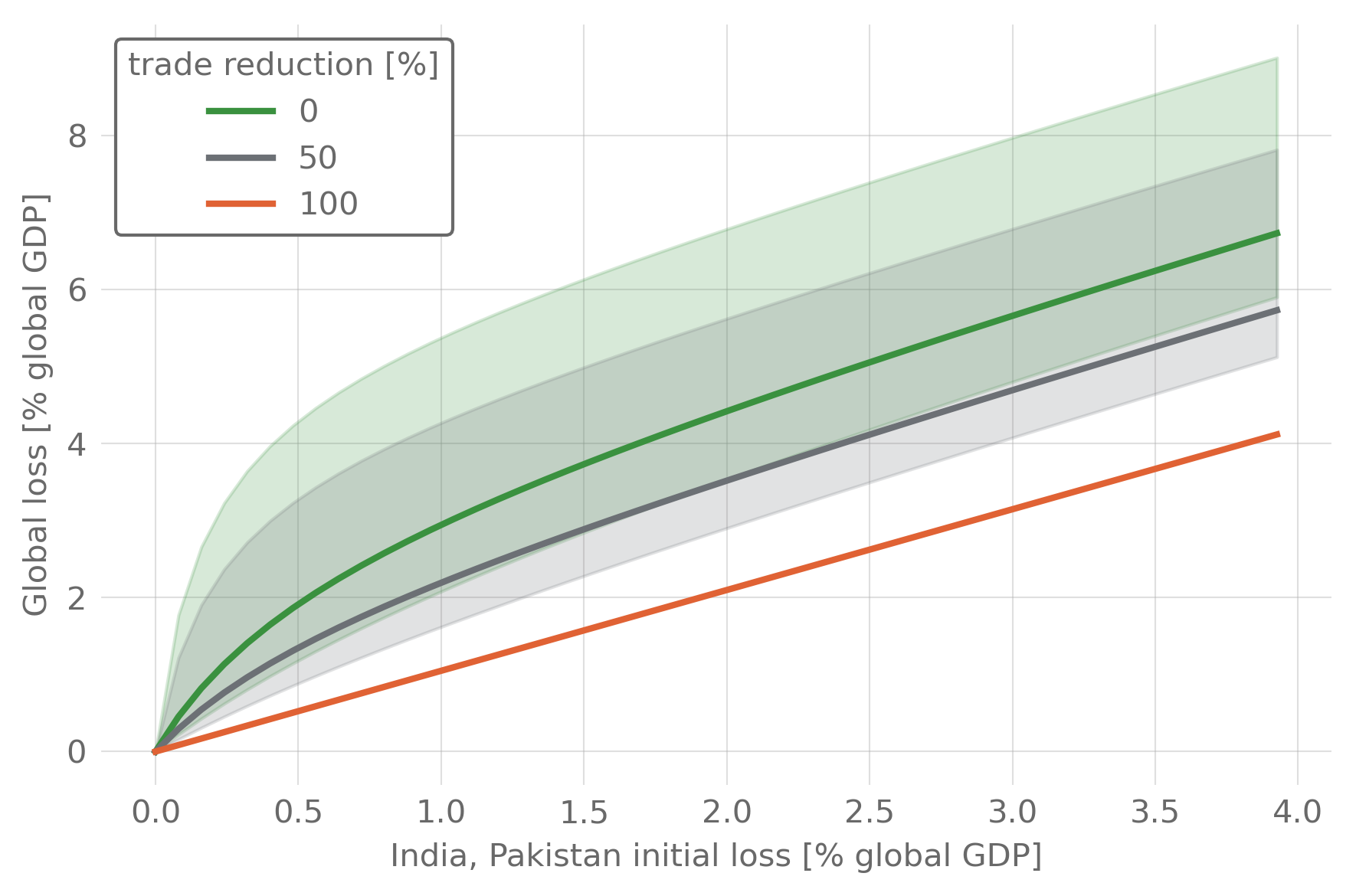}
    \caption{India-Pakistan conflict predictions. On the X-axis, we show India and Pakistan's losses as a percentage of the global GDP, and on the Y-axis, the total global loss (also as a percentage of the global GDP). Results are $50\%$ prediction intervals except for the orange line, in which the only losses are India’s and Pakistan’s. Colours indicate the amount of trade diverted from India and Pakistan to other countries}
    \label{fig:india_pak_global}
\end{figure}

\begin{figure}
    \centering
    \includegraphics[width=\linewidth]{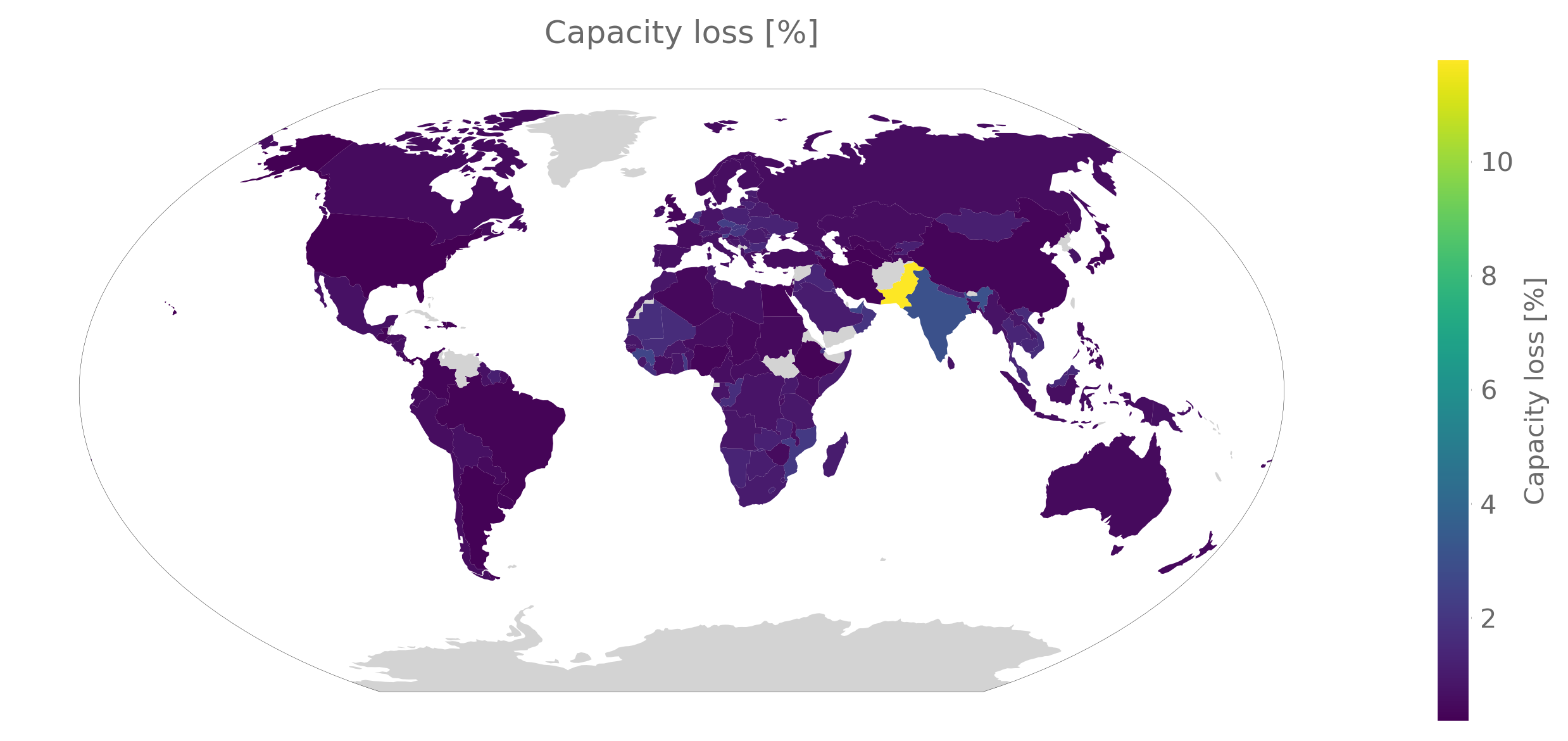}
    \caption{World map indicating relative losses (median prediction) for each country (as long as we have sufficient data to do so) in the example scenario of $2.7\%$ and $11.6\%$ initial capacity losses in India and Pakistan, respectively}
    \label{fig:map}
\end{figure}

\bibliography{ms}